\documentclass[]{pasj01}
\draft

\begin{document} 

\title{Detailed analysis of the poorly studied northern open cluster NGC 1348 using multi-color photometry and GAIA~EDR3 astrometry.}


\author{D. Bisht\altaffilmark{1}$^{*}$, Qingfeng Zhu \altaffilmark{1}, W. H. Elsanhoury\altaffilmark{2,3},
Devesh P. Sariya\altaffilmark{4}, Geeta Rangwal\altaffilmark{5}, R. K. S. Yadav\altaffilmark{6}, Alok Durgapal\altaffilmark{5},
Ing-Guey Jiang\altaffilmark{4}}
\altaffiltext{1}{Key Laboratory for Researches in Galaxies and Cosmology, University of Science and Technology of China, Chinese
                 Academy of Sciences, Hefei, Anhui, 230026, China}
\altaffiltext{2}{Astronomy Department, National Research Institute of Astronomy and Geophysics (NRIAG), 11421, Helwan, Cario, Egypt
           (Affiliation ID: 60030681)}
\altaffiltext{3}{Physics Department, Faculty of Science and Arts, Northern Border University, Turaif Branch, Saudi Arabia}
\altaffiltext{4}{Department of Physics and Institute of Astronomy, National Tsing Hua University, Hsin-Chu, Taiwan.}
\altaffiltext{5}{Center of Advanced Study, Department of Physics, D. S. B. Campus, Kumaun University Nainital 263002, India.}
\altaffiltext{5}{Aryabhatta Research Institute of Observational Sciences, Manora Peak, Nainital 263002, India.}
\email{dbisht@ustc.edu.cn}



\maketitle

\begin{abstract}
The membership determination for open clusters in noisy environments of the Milky Way is still an open problem.
In this paper, our main aim is provide the membership probability of stars using proper motions and parallax values of stars
using Gaia EDR3 astrometry. Apart from the Gaia astrometry, we have also used other photometric data sets like UKIDSS, WISE,
APASS and Pan-STARRS1 in order to understand cluster properties from optical to mid-infrared regions. We selected 438
likely members with membership probability higher than $50\%$ and G$\le$20 mag. We obtained the mean value of proper motion as
$\mu_{x}=1.27\pm0.001$ and $\mu_{y}=-0.73\pm0.002$ mas yr$^{-1}$. The cluster's radius is determined as 7.5 arcmin (5.67 pc)
using radial density profile. Our analysis suggests that NGC 1348 is located at a distance of $2.6\pm0.05$ kpc. The mass
function slope is found to be $1.30\pm0.18$ in the mass range 1.0$-$4.1 $M_\odot$, which is in fair agreement with
Salpeter's value within the 1$\sigma$ uncertainty. The present study validates that NGC 1348 is a dynamically relaxed
cluster. We computed the apex coordinates $(A, D)$ for NGC 1348 as 
$(A_\circ, D_\circ)$ = $(-23^{\textrm{o}}.815\pm 0^{\textrm{o}}.135$, $-22^{\textrm{o}}.228\pm 0^{\textrm{o}}.105)$.
In addition, calculations of the velocity ellipsoid parameters (VEPs), matrix elements $\mu_{ij}$, direction cosines
($l_j$, $m_j$, $n_j$) and the Galactic longitude of the vertex have been also conducted in this analysis.
\end{abstract}


\section{Introduction}
\label{sec:intro}

Open clusters (OCs) have been used to find out the spiral arm structure and evolution of the Galactic disk (Trumpler 1930,
Janes \& Adler 1982, Carraro et al. 1998, Chen et al. 2003, Piskunov et al. 2006, Moraux 2016). 
Due to their location in the disc, open clusters are highly contaminated by the non-member 
stars. Data from the Gaia mission is very helpful in this direction. In continuation
of its previous two data releases the (early) Third data release (hereafter
EDR3; Gaia Collaboration et al. 2020) was made public on 3$^{rd}$ December 2020.
This catalog consists of the central coordinates, proper motions in right ascension and declination and parallaxes $(\alpha, \delta,
\mu_{\alpha}cos\delta, \mu_{\delta}, \pi)$ for more than 1.46 billion sources. Gaia~EDR3 has enabled a breakthrough in OC studies
because it provides accurate information of proper motion and parallaxes for a large number of stars. Cantat-Gaudin et al. (2018) reported
membership probabilities for 1229 OCs with 60 previously unknown clusters based on the Gaia data. One of the most important outcome from the Gaia
data is that we can detect many new OCs. Sim et al. (2019), Liu \& Pang (2019) and Castro-Ginard et al. (2020) identified 207, 76 and 582
new OCs in the Galactic disk. In this paper, our main goal is to perform a detailed analysis of NGC 1348 using Gaia data. The open
cluster NGC 1348 ($\alpha_{2000} = 03^{h}34^{m}06^{s}$, $\delta_{2000}=51^{\circ} 24^{\prime} 30^{\prime\prime}$;
$l$=146$^\circ$.969, $b$=-3$^\circ$.709) is located in the second Galactic quadrant. Carraro (2002) analyzed this object
using CCD UBVI data. He found that NGC 1348 is a significantly reddened cluster $(E(B-V)=0.85)$, lies at a distance $1.9\pm0.5$ kpc and
has an age greater than 50 Myr.


Open clusters contain a spectrum of stellar masses (from very low to high mass stars)
formed from the same molecular cloud. This makes them the ideal objects to study the
initial mass function (IMF).
Many authors have studied IMF in open
clusters (Durgapal \& Pandey 2001, Phelps \& Janes 1993, Piatti et al. 2002, Piskunov et al. 2004, Scalo et al. 1998, Sung and Bessell 2004,
Yadav \& Sagar 2002, 2004, and Bisht et al. 2017 \& 2019). The universality of IMF is still a matter of intense debate
(Elmegreen 2000; Larson 1999; Marks et al. 2012; Dib 2014; Dib, Schmeja \& Hony 2017). 
One of the motives of the present analysis is to gather information of the IMF to
understand the star formation history in NGC 1348. The mass segregation studies
in the OCs provide information about the distribution of stars according to their masses
within the cluster region. The information contained in both
the mass distribution of stars and their spatial distribution can help us understand the process of star formation. We also investigate the
orbits of stars in NGC 1348 as these are very useful to constrain the role of external tidal forces and help us better understand the
dynamical evolution of the cluster.

Virtually, members of a star cluster appear as coherent and mutually associated moving groups of stars sharing similar properties like
distance, kinematics, chemical composition, and age as well as the line of sight velocity (radial velocity). 
The determination of the convergent point coordinates $(A_{\circ} , D_{\circ})$ at which the stars of the cluster seem to be merging
(i.e. apex) is an important parameter in the kinematical and physical examination (Wayman 1965, Hanson 1975, Eggen 1984,
Gunn et al. 1988). To determine the apex, numerous techniques are available in the literature, like i) classical convergent
point method, ii) the AD-chart method, and iii) convergent point search method (CPSM; Galli et al. 2012).
The convergent point method is a classical method which still attracts the 
interest of many working groups. 
This method allows selecting stars based on the parallelism of the proper motion components. 
It was further developed and discussed by Smart (1938), Brown (1950), \& Jones (1971). 
Based on the works by Jones (1971) and de Bruijne (1999), Galli et al. (2012) 
presented the CPSM which also uses the proper motion data. 
The CPSM represents the stellar proper motions by great circles over the celestial sphere 
and visualizes their intersections as the convergent point of the moving group. 
However, for a complete picture of a star’s space motion, 
both proper motions and radial velocities are required. 
Thus, one can identify  the stellar groupings with a common movement in space 
via the AD-chart method based. 
This method that takes into account the 
individual stellar apexes is discussed by Chupina et al. (2001, 2006).
In this work, we adopted the AD-chart method (stellar apex method) for NGC 1348. 
We used the distribution of individual apexes of cluster members in the equatorial coordinate system. 
Also, some kinematical parameters and velocity ellipsoid parameters (VEPs) 
are presented here with the computational algorithm presented in our previous papers 
(Elsanhoury et al. 2015, 2018, Postnikova et al. 2020, Bisht et al. 2020).

The structure of the article is as follows. 
A brief description of the different data sets used here is given in Section 2. 
In Section 3, we performed the
study of proper motion and selected the cluster member stars. 
The structural properties the cluster and derivation of its fundamental parameters are explained
in Section 4. 
Section 5 deals with the study of Luminosity and mass function 
while the mass segregation is described in Section 6
along with dynamical and kinematical analysis of the cluster. 
We conclude the present work in Section 7.

\begin{figure}
\begin{center}
\includegraphics[width=12.5cm, height=12.5cm]{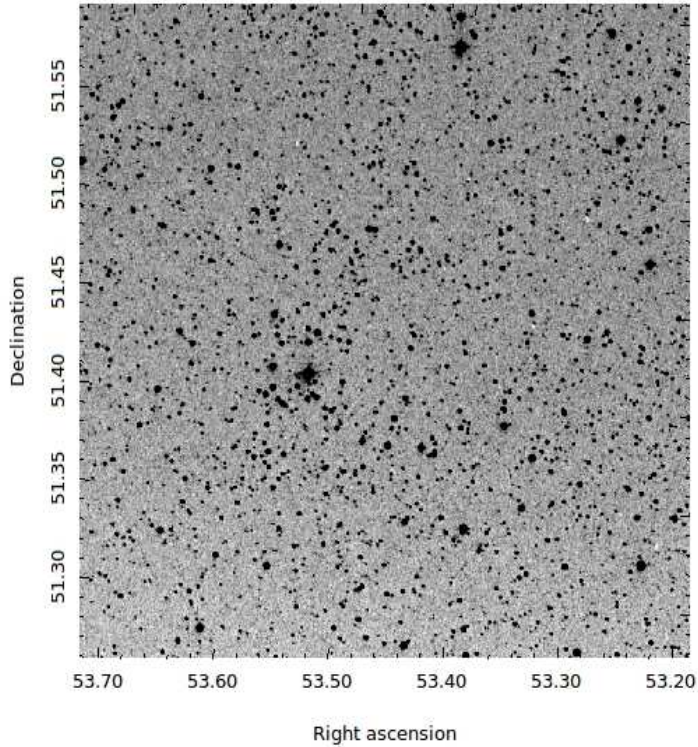}
\caption{The identification map of NGC 1348 taken from the DSS.}
\label{id}
\end{center}
\end{figure}

\begin{figure}
\centering
\includegraphics[width=10.5cm,height=12.5cm]{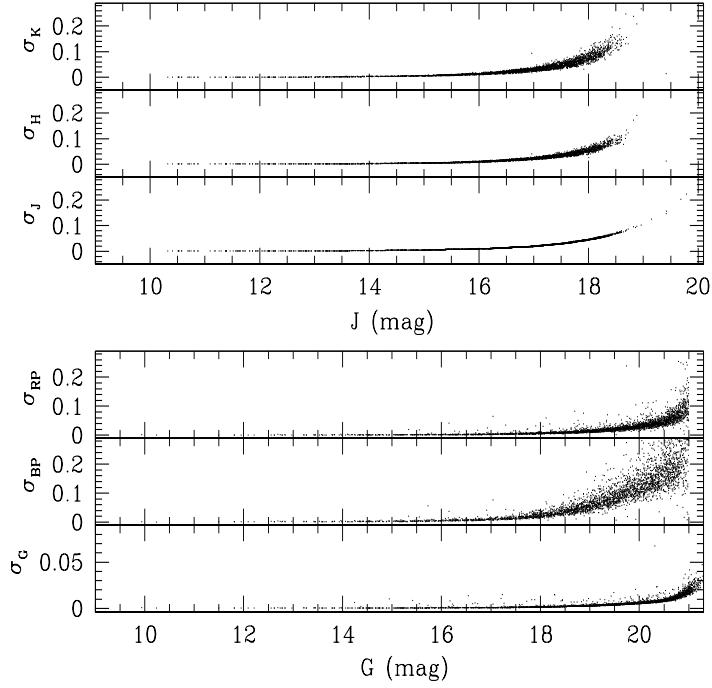}
\caption{Photometric errors in the $J$, $H$ and $K$ magnitudes against $J$ magnitude (upper panels). 
Photometric errors in the Gaia pass bands $G$, $G_{BP}$ and $G_{RP}$ against $G$ magnitude (lower panels).}
\label{error}
\end{figure}

\section{Data}

We extracted photometric data of the cluster within a 10 arcmin radius from the 
APASS, Pan-STARRS1, UKIDSS and WISE along with astrometric data from GAIA EDR3. 
The main purpose is to take different photometric surveys' data 
to check the extinction law towards the open cluster NGC 1348 from the optical to
the mid-infrared. 
After cross-matching all these catalogs, the fundamental parameters, mass function, 
Galactic orbits and kinematics have been studied in the current paper. 
The identification map shown in the Fig.~\ref{id} is taken from the Digitized Sky Survey (DSS). 
The descriptions of the above mentioned data sets are as following:

\subsection{\bf GAIA EDR3}

\begin{figure}
\centering
\includegraphics[width=10.5cm,height=12.5cm]{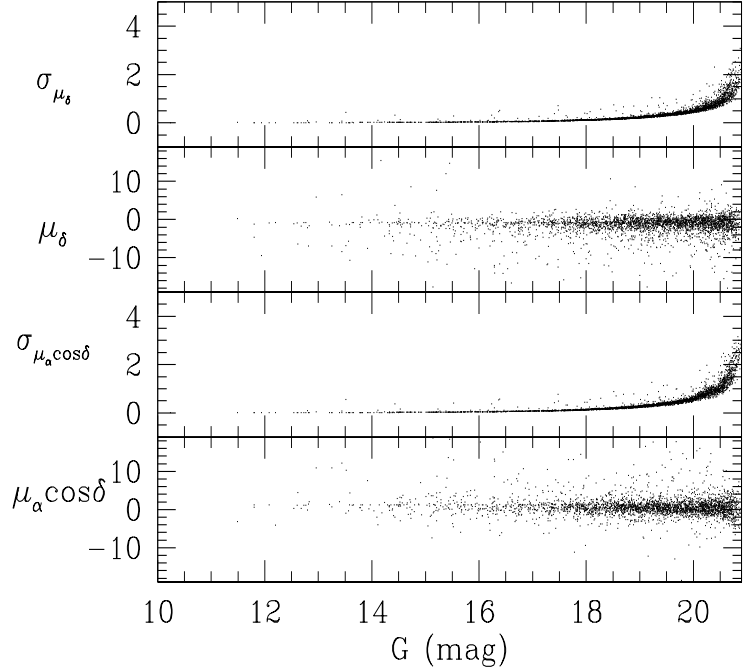}
\caption{Plot of Proper motions and their errors versus $G$ magnitude. The unit of proper motions and their errors is mas/yr.}
\label{error_proper}
\end{figure}

We have used GAIA~EDR3 (Gaia Collaboration et al. 2020) data for the astrometric investigation of NGC 1348. This data
consists of five quantities, which are position coordinates
, parallaxes and proper motions in two directions having a limiting magnitude of $G=21$ mag. We have plotted the errors in the three photometric bands
($G$, $G_{BP}$ and $G_{RP}$) along with their $G$ magnitudes as shown in the three bottom panels of Fig \ref{error}.
For the sources having G$\le$ 15 mag, the uncertainties in parallax are $\sim$ 
0.02-0.03 mas while for the sources with G$\le$ 17 mag, it is $\sim$ 0.07.
In Fig. \ref{error_proper}, we plotted the proper motion and their corresponding
errors as a function of $G$ magnitude. This figure shows that the maximum
error in proper motion components is $\sim 0.4$ mas/yr upto $G\sim20$ mag.


\subsection{\bf UKIDSS}

The UKIRT Infrared Deep Sky Survey (UKIDSS; Lawrence et al. 2007) is a deep large scale infrared survey organized with
the Wide Field Camera (WFCAM; Casali et al. 2007) on UKIRT. The UKIDSS GCS DR9 covers $\sim$ 36 square degrees observed
in five passbands ($Z$, $Y$, $J$, $H$, $K$; Hewett et al. 2006).

\subsection{\bf WISE}

The WISE database contains photometric magnitudes of stars in the mid-IR bands. The effective wavelength of these bands are $3.35 \mu m (W1)$,
$4.60 \mu m (W2)$, $11.56 \mu m (W3)$ and $22.09 \mu m (W4)$ (Wright et al. 2010). We have extracted data from the ALLWISE source
catalog for NGC 1348.

\subsection{\bf APASS}

The American Association of Variable Star Observers (AAVSO) Photometric All-Sky Survey (APASS) is cataloged
in five filters: B, V (Landolt) and $g^{\prime}$, $r^{\prime}$, $i^{\prime}$, with $V$ band magnitude range
from 7 to 17 mag (Heden \& Munari 2014). The DR9 catalog covers about $99\%$ of the sky (Heden et al. 2016).
From here, we have used data in $B$ and $V$ bands for NGC 1348. 

\subsection{\bf Pan-STARRS1}

The Pan-STARRS1 survey (Hodapp et al. 2004) provides data in five broad-band filters, $g$, $r$, $i$, $z$, $y$, screening
from 400 nm to 1 $\mu$m (Stubbs et al. 2010). 
These data have a mean 5-$\sigma$ point
source limiting sensitivities as 23.3, 23.2, 23.1, 22.3, and 21.4 mag in $g$, 
$r$, $i$, $z$, and $y$ bands respectively (Chambers et al. 2016). The filters
have an effective wavelengths of 481, 617, 752, 866, and 962 nm, respectively (Schlafly et al. 2012; Tonry et al. 2012). 

\section{Mean Proper motion and Membership probability of stars}

\begin{figure*}
\centering
\includegraphics[width=12.5cm, height=12.5cm]{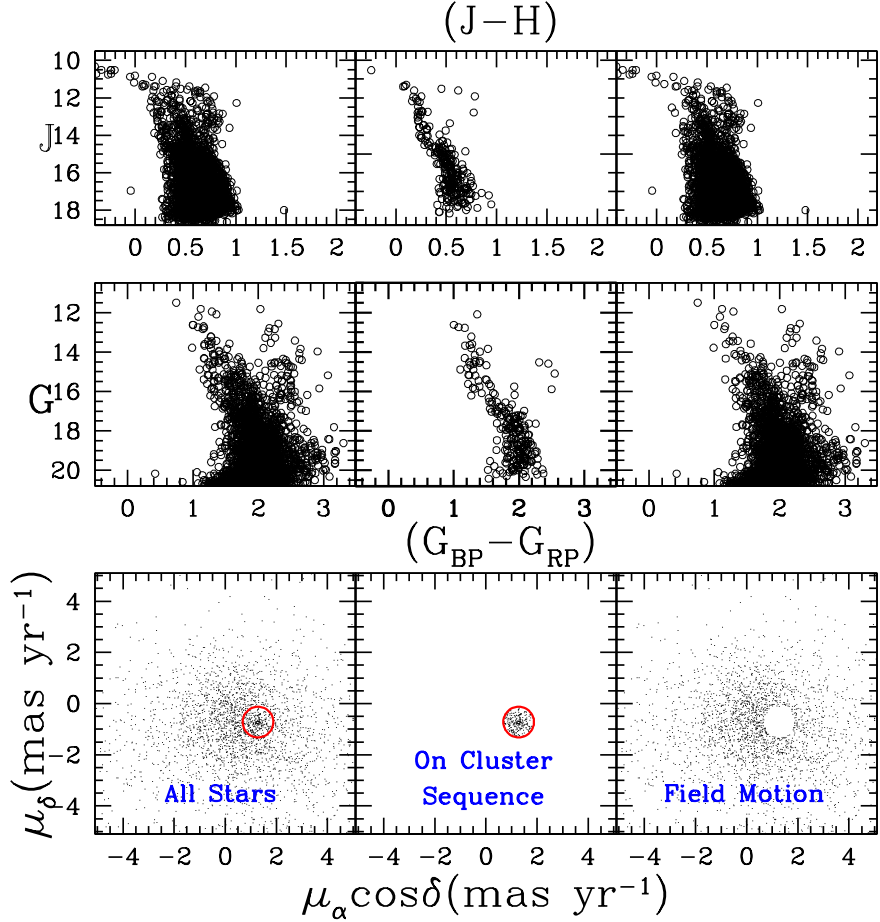}
\caption{(Bottom panels) Proper-motion vector point diagrams (VPDs) for NGC 1348. (Top panels) $J$ versus $(J-H)$ color 
magnitude diagrams. (Middle panels) $G$ versus $(G_{BP}-G_{RP})$ color magnitude diagrams. (Left panel) The entire
sample. (Center) Stars within the circle of 0.6~ mas~ yr$^{-1}$ radius centered around the mean proper motion.
(Right) Probable background/foreground field stars in the direction of the cluster. 
All these plots show only the stars
with PM error smaller than 0.5~mas~ yr$^{-1}$ in each coordinate.}
\label{pm_dist}
\end{figure*}

We plotted a diagram between the Proper motions (PMs) 
($\mu_{\alpha} cos{\delta}$, $\mu{\delta}$) 
which is called Vector Point Diagrams (VPDs) and shown
in the bottom panels of Fig. \ref{pm_dist}. 
The top and middle panels shows that the corresponding $G$ versus $(G_{BP}-G_{RP})$
and $J$ versus ($J-H$) color magnitude diagrams (CMDs). 
The left panel shows all
stars within a radius of of 10 arcmin around the cluster center, while the 
middle and right panels show the probable cluster members having similar motion 
in the sky and non-member stars, respectively. The selection
of circle's radius as 0.6 mas/yr in VPD is a compromise between losing stars with poor PMs and the contamination of
field stars. The CMD of the selected probable cluster members is shown in the upper-middle panels in Fig. \ref{pm_dist}. The main
sequence of the cluster is clearly separated from the non members.

For the mean proper motion estimation, we consider the only probable cluster 
members on the basis of clusters VPD and CMD as shown
in Fig. \ref{pm_dist}. By using weighted mean method, we found the mean-proper motion of NGC 1348 as $1.27\pm0.001$ and
$-0.73\pm0.002$ mas yr$^{-1}$ in RA and DEC directions, respectively.

In this paper, we used the method described by Balaguer-N\'{u}\~{n}ez et al. (1998) 
by using Gaia EDR3 catalog data for NGC 1348
to estimate the membership probability of stars. This method has been used for several clusters by various authors
(Yadav et al. 2013; Sariya et al. 2021a, 2021b; Bisht et al. 2020). Recently we have adopted the above membership
probability method for few OCs using Gaia EDR3 data ( Bisht et al. 2021a, Bisht et al. 2021b). We used stars with PM
errors $\le$ 0.5 mas/yr to express cluster and field star distributions. A group of stars is found at
$\mu_{xc}$=1.27 mas~yr$^{-1}$, $\mu_{yc}$=$-$0.73 mas/yr. Considering a distance of 2.6 kpc and radial velocity
dispersion of 1 km $s^{-1}$ for open star clusters (Girard et al. 1989), the expected dispersion ($\sigma_c$) in
PMs would be 0.08 mas/yr. For the non-members, we obtained ($\mu_{xf}$, $\mu_{yf}$) = ($-$1.0, $-$1.7) mas/yr and
($\sigma_{xf}$, $\sigma_{yf}$) = (3.9, 2.6) mas/yr.

\begin{figure}
\begin{center}
\centering
\hbox{
\includegraphics[width=7.5cm, height=7.5cm]{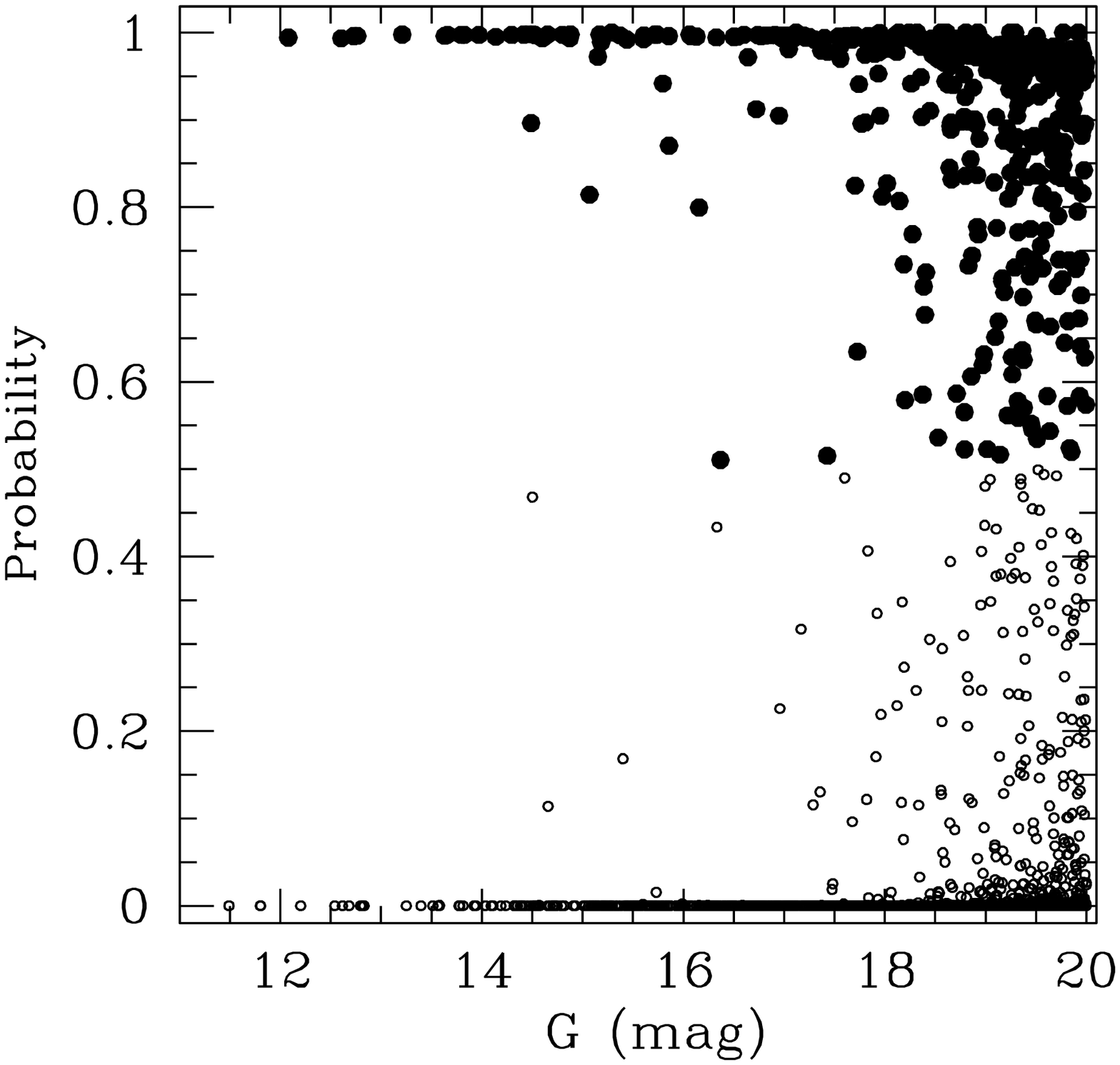}
\includegraphics[width=7.5cm, height=7.5cm]{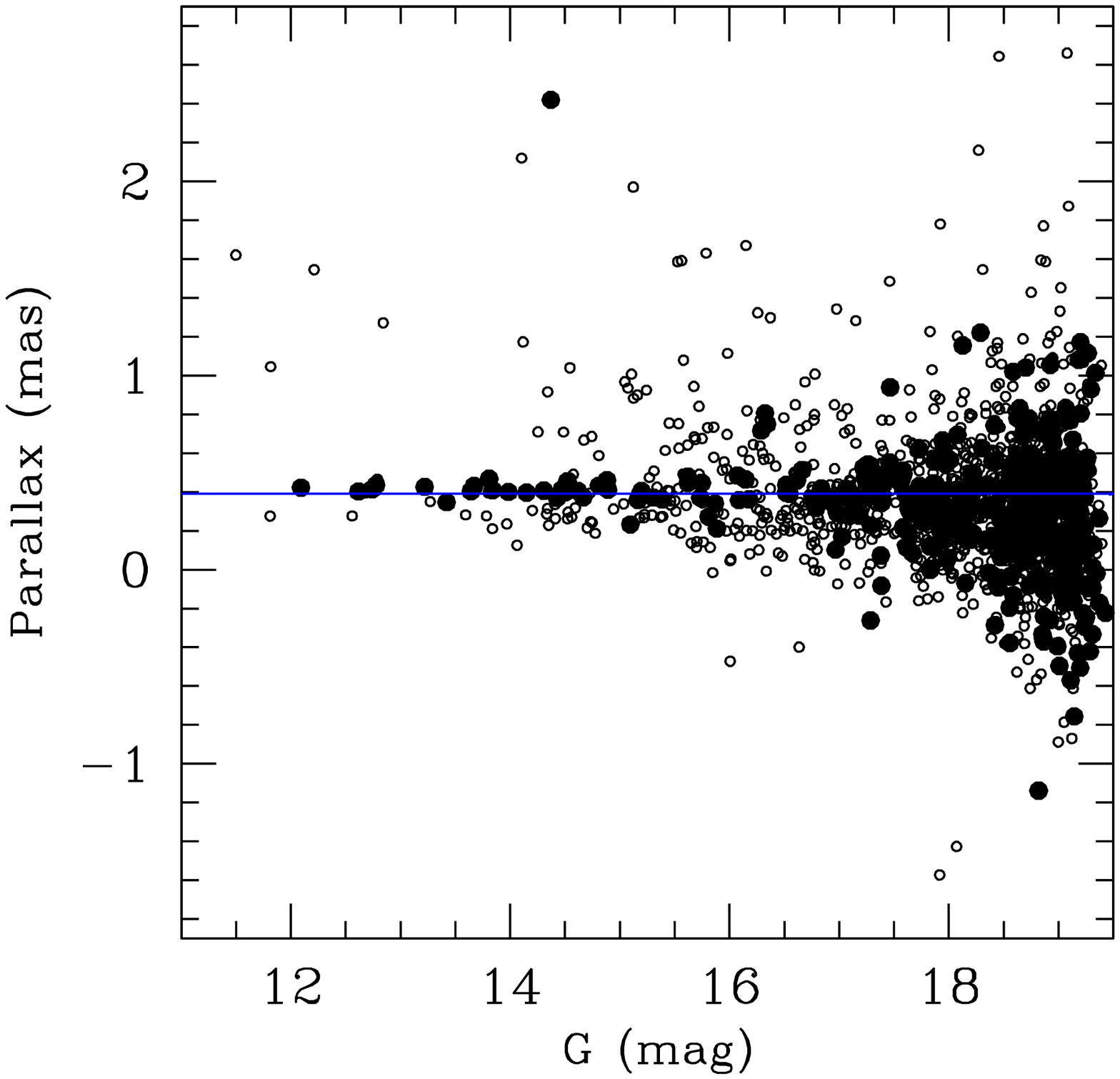}
}
\caption{(Left panel) The cluster membership probabilities plotted with G magnitude. (Right panel) Cluster
parallax with G magnitude. Solid black dots are probable cluster members with membership probability higher than $50\%$.
}
\label{membership}
\end{center}
\end{figure}

\begin{figure}
\begin{center}
\centering
\includegraphics[width=12.5cm, height=12.5cm]{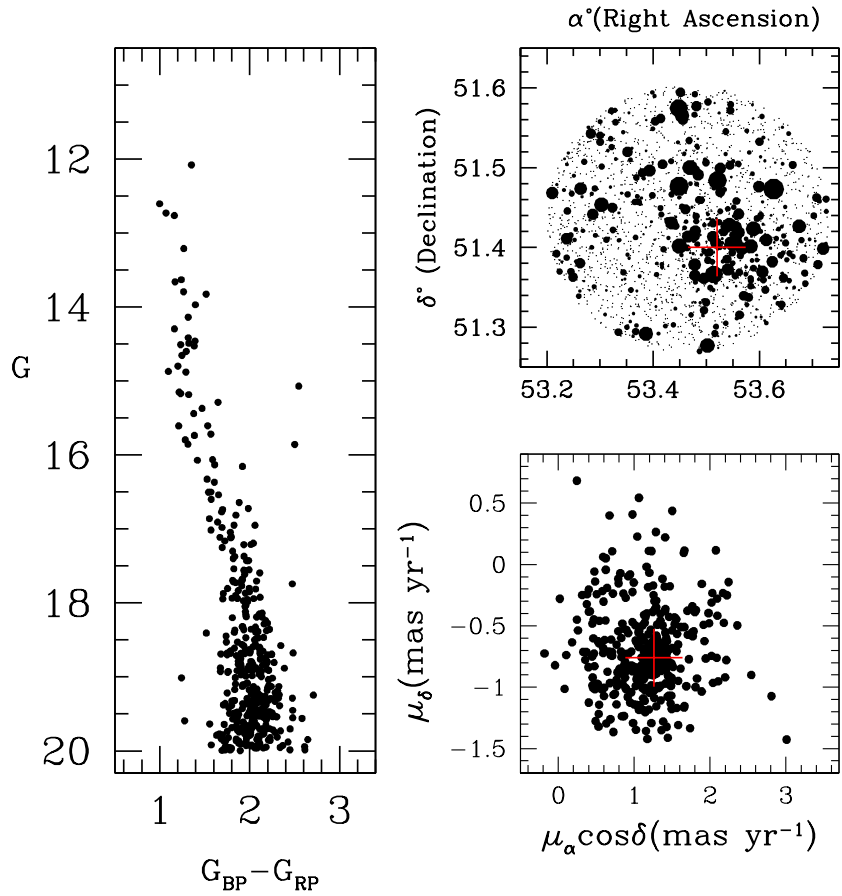}
\caption{($G, G_{BP}-G_{RP}$) CMD, identification chart and proper motion distribution of member stars with membership
probability higher than $50\%$. The plus sign indicates the cluster center.
}
\label{new_membership}
\end{center}
\end{figure}

Based on the above method, 438 stars are selected as member stars with membership probability higher than $50\%$
and $G\le20$ mag. In the left panel of Fig.~\ref{membership}, we plotted membership probability versus $G$ magnitude.
In this figure, we can see a clear separation of the cluster and the field stars. In the right panel of this figure, we plotted $G$
magnitude versus parallax of stars. The most probable cluster members with high membership probability  $(\ge 50\%)$ are
shown by red dots in Fig.~\ref{membership}. we have plotted $G$ versus ($G_{BP}-G_{RP}$) CMD, the identification
chart and proper motion distribution using stars with membership probability higher than $50\%$ in Fig. \ref{new_membership}.
The Cantat-Gaudin et al. (2018) catalog reports membership probabilities
for the stars of this cluster but only up to 18 mag in $G$ band. 
Here, We provide the most probable cluster members up to 20 mag 
in $G$ band which is fainter than Cantat-Gaudin et al. (2018).


\section{Clusters Structure, extinction law and fundamental parameters evaluation}

\subsection{Cluster radius and radial stellar surface density}

To estimate the cluster's center, the weighted mean of the positions of all stars has been considered by von Hoerner (1960, 1963). 
The center can be estimated by fitting a Gaussian function to the star's distribution and taking the center to be the point
of maximum number density. We adopted this  method to find the central coordinates of NGC 1348. This method has been
described by Bisht et al. (2020). The central coordinates are found as $\alpha = 53.51\pm0.03$ deg ($3^{h} 34^{m} 2.3^{s}$)
and $\delta = 51.41\pm0.02$ deg ($51^{\circ} 24^{\prime} 36^{\prime\prime}$) which are in good agreement with the values given
in Dias et al. (2002).

\begin{figure}
\hspace{3cm}\includegraphics[width=10.5cm, height=10.5cm]{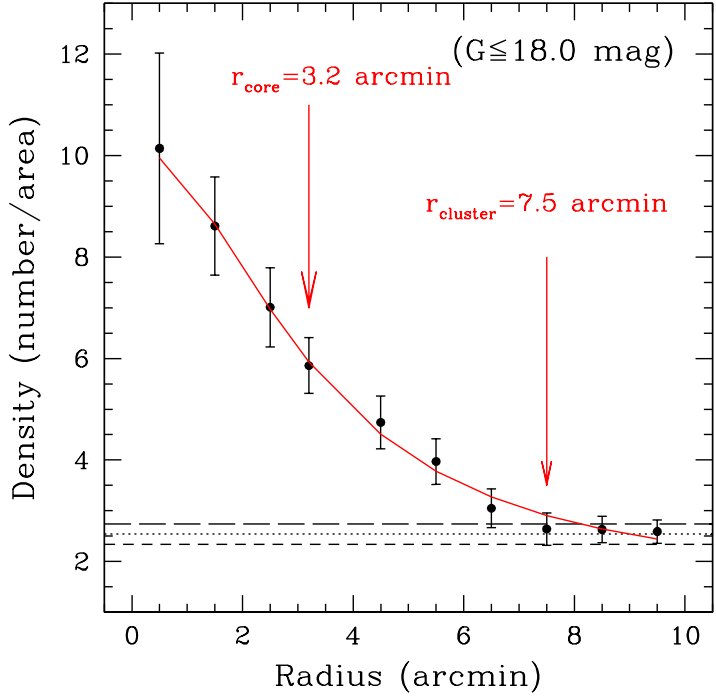}
\caption{Surface density distribution of the cluster NGC 1348 using GAIA~EDR3 $G$ band data. Errors are determined from
sampling  statistics (=$\frac{1}{\sqrt{N}}$ where $N$ is the number of cluster members used in the density estimation at
that point). The smooth line represents the fitted profile of King (1962)  whereas the dotted line shows the background density
level. Long and short dash lines represent the errors in background density.} 
\label{dens}
\end{figure}

After center estimation, the next step is to construct a radial density profile 
(RDP), for which, we have drawn many concentric rings
around the cluster center using the above estimated values
of center coordinates. We determined the stellar number density, $\rho_{i}$, in
the $i^{th}$ zone of the cluster by using the relation: $\rho_{i}$ = $\frac{N_{i}}{A_{i}}$, 
where $N_{i}$ is the number of cluster members in the area $A_{i}$ of the $i^{th}$ 
zone. 
By fitting the King (1962) profile in this distribution as shown by a smooth continuous line in Fig. \ref{dens}, we determined
the structural properties of the cluster. The King (1962) profile is given as:\\

~~~~~~~~~~~~~~~~~~~~~~~${\bf f(r) = f_{bg}+\frac{f_{0}}{1+(r/r_{c})^2}}$\\

where $r_{c}$, $f_{0}$, and $f_{bg}$ are the core radius, central density, and the background density level, respectively.

We have shown background density level with errors using dotted lines in
Fig. \ref{dens}. At $r\sim$ 7.5$^{\prime}$ cluster stars get merged with 
the non-member stars, as shown clearly in Fig. \ref{dens}. 
Hence, we considered 7.5$^{\prime}$ as the cluster radius.
The error bars are calculated using the Poisson statistics error in each shell as $P_{err}=\frac{1}{\sqrt{N}}$. By fitting the
King model to the cluster density profile, the structural parameters are found as: $f_{b}$=2.54 star/arcmin$^{2}$,
$f_{0}$=10.15 star/arcmin$^{2}$ and $r_{c}$=3.2 arcmin. We obtained the density contrast parameter ($\delta_{c}$) using the
formula described by Bisht et al. (2020), which indicates that NGC 1348 is a sparse cluster. The tidal radius of clusters
is normally influenced by the effects of Galactic tidal fields and later by internal relaxation dynamical evolution of clusters
(Allen \& Martos 1988). 
To calculate the tidal radius of NGC 1348, we used the formula derived by
Bertin \& Varri (2008) as:

~~~~~~~~~~~~~~~~~~~$r_{t}=(\frac{GM_{cl}}{\omega^{2} \nu})^{1/3}$ \\

where $\omega$ and $\nu$  are   \\

$\omega= (d\Phi_{G}(R)/dR)_{R_{gc}}/R_{gc})^{1/2}$  \\
$\nu=4-\kappa^{2}/\omega^{2}$     \\

where $\kappa$ is \\

$\kappa=(3\omega^{2}+(d^{2}\Phi_{G}(R)/dR^{2})_{R_{gc}})^{1/2}$  \\

here $\Phi_{G}$ is Galactic potential, $M_{cl}$
mass of the cluster, $R_{gc}$ is the Galactocentric
distance of the cluster, $\omega$ is the orbital frequency, $\kappa$ is
epicyclic frequency and $\nu$ is a positive constant. We used the Galactic
potentials discussed in section 5. The value of the Galactocentric distance is 
taken from Table~1 and mass of the cluster is taken from section 6.
In this manner, the tidal radius of the cluster is calculated as 9.2 pc.

\subsection{Optical to mid-infrared extinction law}

\begin{figure*}
\begin{center}
\centering
\includegraphics[width=10.5cm, height=10.5cm]{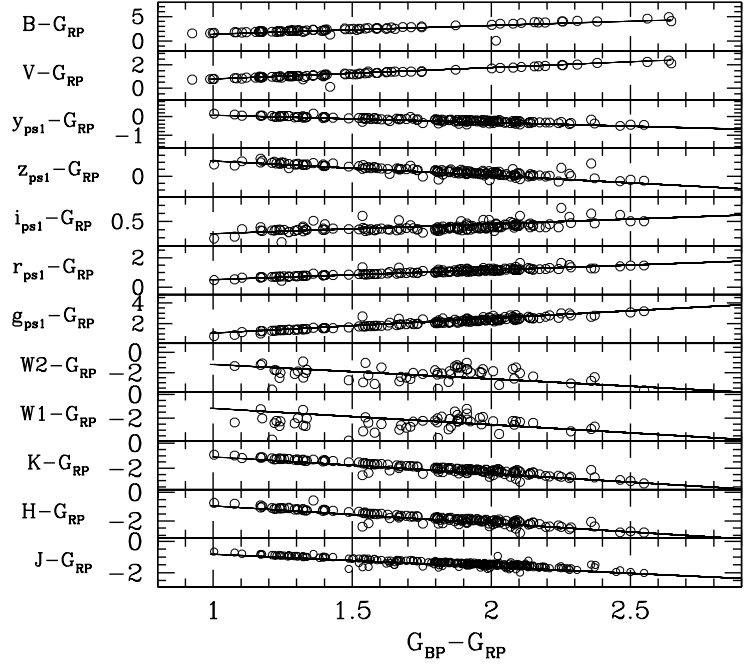}
\caption{The $(\lambda-G_{RP})/(G_{BP}-G_{RP})$ TCDs for the stars selected from VPD of NGC 1348.
The continuous blue lines represent the slope determined through the least-squares linear fit.
}
\label{cc_gaia}
\end{center}
\end{figure*}

We have matched the multi-wavelength photometric data with Gaia astrometry to study the extinction law in various wavebands
for NGC 1348. We plotted various $(\lambda-G_{RP})/(G_{BP}-G_{RP})$ two-color diagrams (TCDs) as shown in Fig. \ref{cc_gaia}. Here,
$\lambda$ represent the filters other than $G_{RP}$. A linear fit was executed in all TCDs to find the slope, which are listed in
Table \ref{gaia_slope}. These values of slopes are in fair agreement with the value described by Wang and Chen (2019).
The value of total-to-selective absorption ratios $R_{cluster}$ 
in the range of $\sim$ 2.9-3.3 for different pass bands demonstrates that the
reddening law is normal towards the cluster region of NGC 1348.

\begin{table*}
\caption{Multi-band color excess ratios in the direction of NGC 1348.
}
\vspace{0.5cm}
\centering
\begin{center}
\small
\begin{tabular}{ccc}
\hline\hline
Band $(\lambda)$ & Effective wavelength  &   $\frac{\lambda-G_{RP}}{G_{BP}-G_{RP}}$ \\ 
\\
\hline\hline
Johnson~ B        &445              &$1.66\pm0.01$\\
Johnson~ V        &551              &$1.05\pm0.02$\\
Pan-STARRS~ g     &481              &$1.42\pm0.02$\\
Pan-STARRS~ r     &617              &$0.73\pm0.03$\\
Pan-STARRS~ i     &752              &$0.15\pm0.03$\\
Pan-STARRS~ z     &866              &$-0.16\pm0.05$\\
Pan-STARRS~ y     &962              &$-0.35\pm0.04$\\
UKIDSS~ J          &1234.5           &$-0.79\pm0.03$\\
UKIDSS~ H          &1639.3           &$-1.21\pm0.03$\\
UKIDSS~ K          &2175.7           &$-1.35\pm0.06$\\
WISE ~W1          &3317.2           &$-1.39\pm0.06$\\
WISE~ W2          &4550.1           &$-1.40\pm0.07$\\
\hline
\end{tabular}
\label{gaia_slope}
\end{center}
\end{table*}

\subsection{Reddening from UKIDSS colors}

\begin{figure}
\begin{center}
\centering
\includegraphics[width=8.5cm, height=8.5cm]{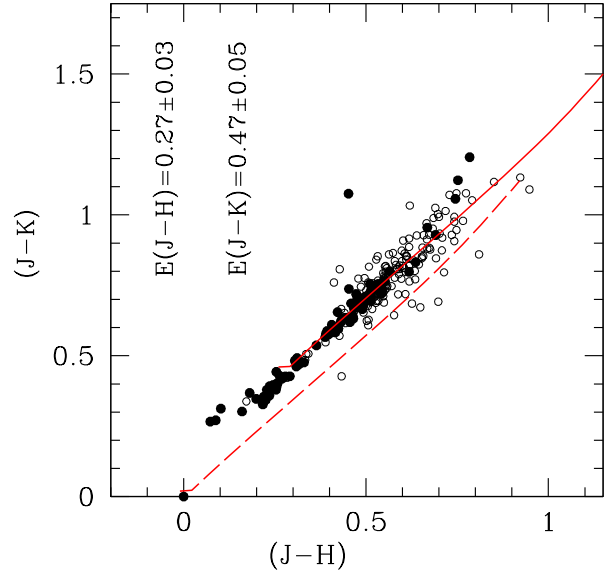}
\caption{The color-color diagram (CCD) for NGC 1348 using the probable cluster members. In this figure, the red
solid line is the ZAMS taken from Caldwell et al. (1993) while the red dotted line is the same ZAMS shifted by the
values given in the text. Solid black dots are the stars matched with Cantat-Gaudin et al. (2018).}
\label{cc}
\end{center}
\end{figure}

The $(J-H)$ versus $(J-K)$ color-color diagram plot has been used to 
obtain the value of interstellar reddening as shown in Fig \ref{cc}. 
The solid line represents the Zero age main sequence (ZAMS) as taken from Caldwell et al. (1993).
The similar ZAMS shown by the dotted line is displaced by $E(J-H) = 0.27\pm0.03$ mag and $E(J-K) = 0.47\pm0.05$ mag.
The color excess ratio ($\frac{E(J-H)}{E(J-K)}$=0.57) 
is showing agreement with the normal value of 0.55 given by Cardelli et al. (1989). 
We obtained the value of interstellar reddening ($E(B-V)$) as 0.88 mag. Our estimated value
is in good agreement with Carraro (2002). Our criterion for reddening estimation is admissible in exceptionally
extended regions.

\subsection{Age, distance and Galactocentric coordinates}

\begin{figure*}
\begin{center}
\centering
\hbox{
\includegraphics[width=8.5cm, height=8.5cm]{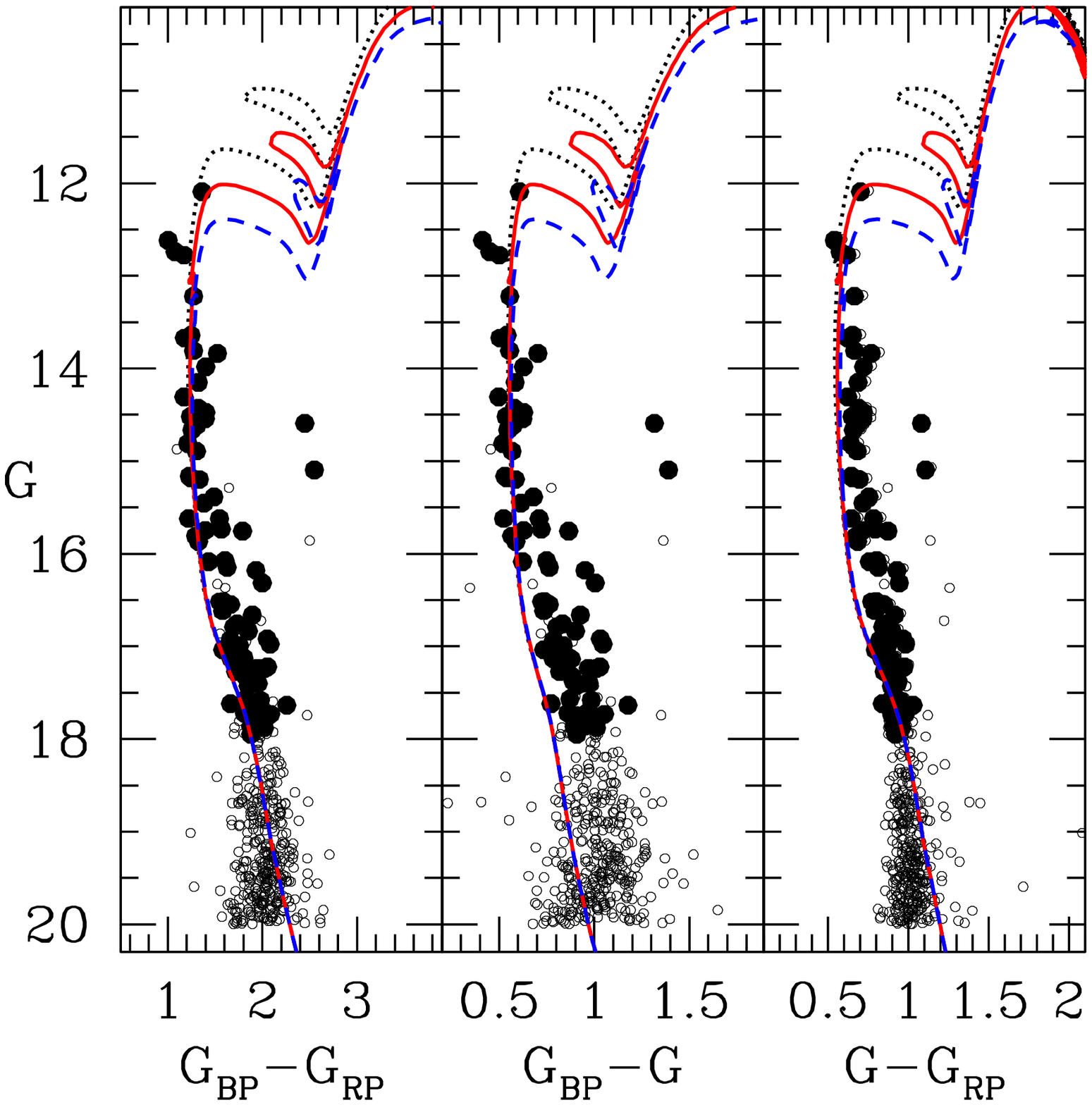}
\includegraphics[width=8.5cm, height=8.5cm]{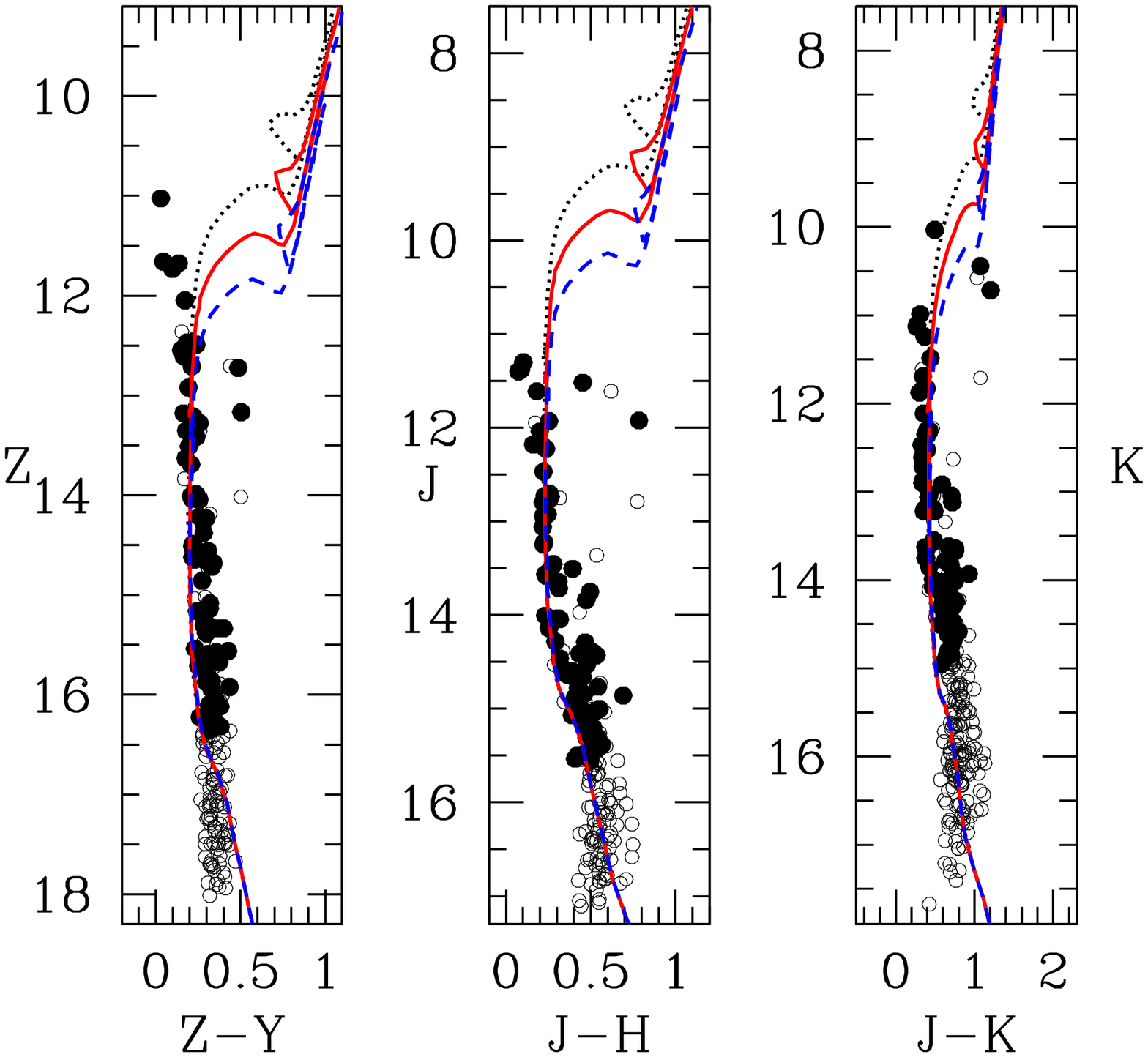}
}
\caption{The $G, (G_{BP}-G_{RP})$, $G, (G_{BP}-G)$, $G, (G-G_{RP})$, $Z, (Z-Y)$, $J, (J-H)$ and $K, (J-K)$ color-magnitude
diagrams of open star cluster NGC 1348. These stars are probable cluster members
The curves are the
isochrones of (log(age)=8.10, 8.20 and 8.30). These isochrones are taken from Marigo et al. (2017). Solid black dots are matched stars
with Cantat-Gaudin et al. (2018).}
\label{cmd}
\end{center}
\end{figure*}

The main fundamental parameters (age, distance, and reddening) have been obtained by fitting the theoretical isochrones
of Marigo et al. (2017) to all the CMDs as shown in Fig. \ref{cmd}. The observed data have been corrected for reddening using
the coefficients ratios $\frac{A_{J}}{A_{V}}$=0.276 and $\frac{A_{H}}{A_{V}}$=0.176, which are taken from Schlegel et al. (1998),
while the ratio $\frac{A_{K_{s}}}{A_{V}}$=0.118 was derived from Dutra et al. (2002).For the Gaia DR2, we have estimated the mean value
of $A_{G}$ and $E(G_{BP}-G_{RP})$ as 1.92 and 0.96 using stars with membership probability higher than $50\%$. Cantat-Gaudin et al. (2018)
catalog contains the membership probabilities of many OCs. In this paper, we have matched our likely members with their catalog and selected
common stars having probability higher than $50\%$. These matched stars have been plotted in the CMDs as shown in Fig. \ref{cmd}.

The isochrones of different ages (log(age)=8.10, 8.20 and 8.30) with $Z=0.008$ have been over plotted on all the CMDs for the cluster
NGC 1348 as shown in Fig \ref{cmd}. The overall fit is satisfactory for log(age)=8.20 (middle isochrone) to the brighter stars,
corresponding to $160\pm40$ Myr. The estimated distance modulus ($(m-M)$=13.80 mag) provides a distance from the Sun that is
$2.4\pm0.10$ kpc.

\subsubsection{Distance of the cluster using parallax angle}

The distance can be estimated using the mean parallax of probable member stars (Luri et al. 2018). By using weighted mean method,
the mean parallax for the cluster is found to be $0.39\pm0.005$ mas. 
Bailer-jones (2015) have shown that the distance estimation 
just by inverting the
parallax is not reliable when there is an associated error.
They described that a correct approach is to obtain the distance values from the parallaxes of stars
through probabilistic analysis which includes a combination of a likelihood (measurements)
and prior (assumption). Bailer-jones (2015) investigated different types of priors and Bailer-Jones
(2018) suggested a exponentially decreasing space density prior in distance $r$.
The prior depends upon a length scale parameter which can be obtained by fitting
a three dimensional model of the Galaxy observed by Gaia and varies smoothly as 
a function of Galactic longitude and latitude. With
the help of this prior, distance of the object can be calculated using a posterior
which is similar as the likelihood (a Gaussian distribution function in parallax)
but a function of distance. This
method gives a pure geometric distance of objects which is independent of
physical properties of interstellar extinction towards an individual star.
Before calculating the distance we corrected the parallax for the offset (-0.017mas)
as suggested by Lindegren et al. (2020) for Gaia EDR3 data set.
Then by adopting the above mentioned method the distance is estimated as $2.6\pm0.05$ kpc.
This value of the cluster's distance is in good agreement with our result obtained from the isochrone fitting method.

\section{Orbits of NGC 1348}

Galactic orbits are very useful to explain the dynamical characteristics of clusters. We derive orbits
and orbital parameters of NGC 1348 using Galactic potential models discussed by Allen \& Santillan (1991). Bajkova \& Bobylev (2016)
and Bobylev et al. (2017) refined the Galactic potential model parameters with the using new observational data for a
distance R$\sim$ 0-200 kpc. The description of these Galactic potential 
models is given by Rangwal et al. (2019).

The input parameters required to calculate orbits of the cluster, such as central coordinates ($\alpha$ and $\delta$),
mean proper motions ($\mu_{\alpha}cos\delta$, $\mu_{\delta}$), parallax angles, age and heliocentric distance ($d_{\odot}$)
have been taken from our investigation in this paper. The radial velocity for this object is not available in the literature.
Average radial velocity for NGC 1348 was calculated by taking the mean of 14 probable cluster members as selected from the
Gaia DR2 catalog. After five iterations, the average radial velocity is found as $-18.71\pm1.60$ km/sec.

The right-handed coordinate system is used to convert equatorial velocity components into Galactic-space velocity components
($U,V,W$), where $U$, $V$ and $W$ are radial, tangential and vertical velocities respectively. Here, the x-axis is taken positive
towards the Galactic-center, the y-axis is along the direction of Galactic rotation and the z-axis is towards the Galactic north pole. Galactic
center is taken at ($17^{h}45^{m}32^{s}.224, -28^{\circ}56^{\prime}10^{\prime\prime}$) 
and the North-Galactic pole is taken to be located at
($12^{h}51^{m}26^{s}.282, 27^{\circ}7^{\prime}42^{\prime\prime}.01$) (Reid \& Brunthaler, 2004). To apply a correction for
Standard Solar Motion and Motion of the Local Standard of Rest (LSR), we used position coordinates of the Sun as ($8.3,0,0.02$)
kpc and its space-velocity components as ($11.1, 12.24, 7.25$) km/s (Schonrich et al. 2010). The transformed parameters in the Galactocentric
coordinate system are listed in Table \ref{inp}.

\begin{figure*}
\begin{center}
\hbox{
\includegraphics[width=6.2cm, height=6.2cm]{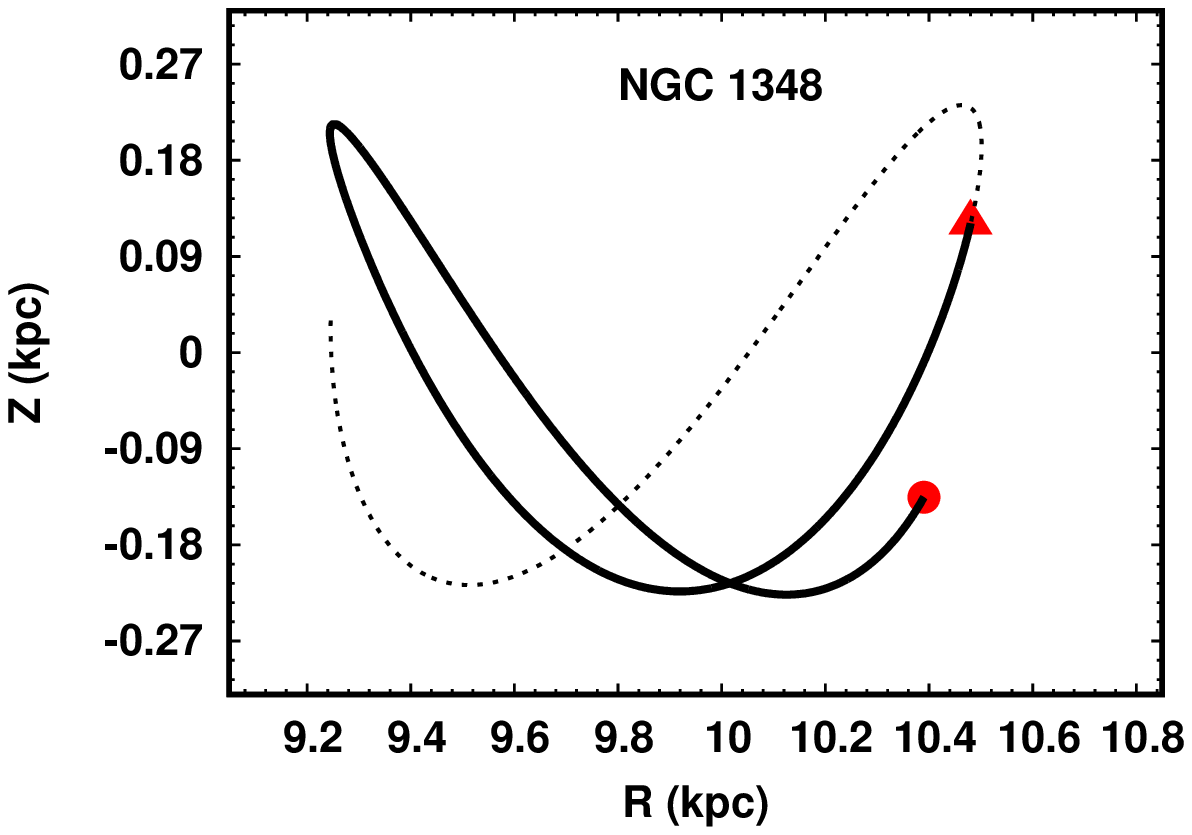}
\includegraphics[width=8.2cm, height=6.2cm]{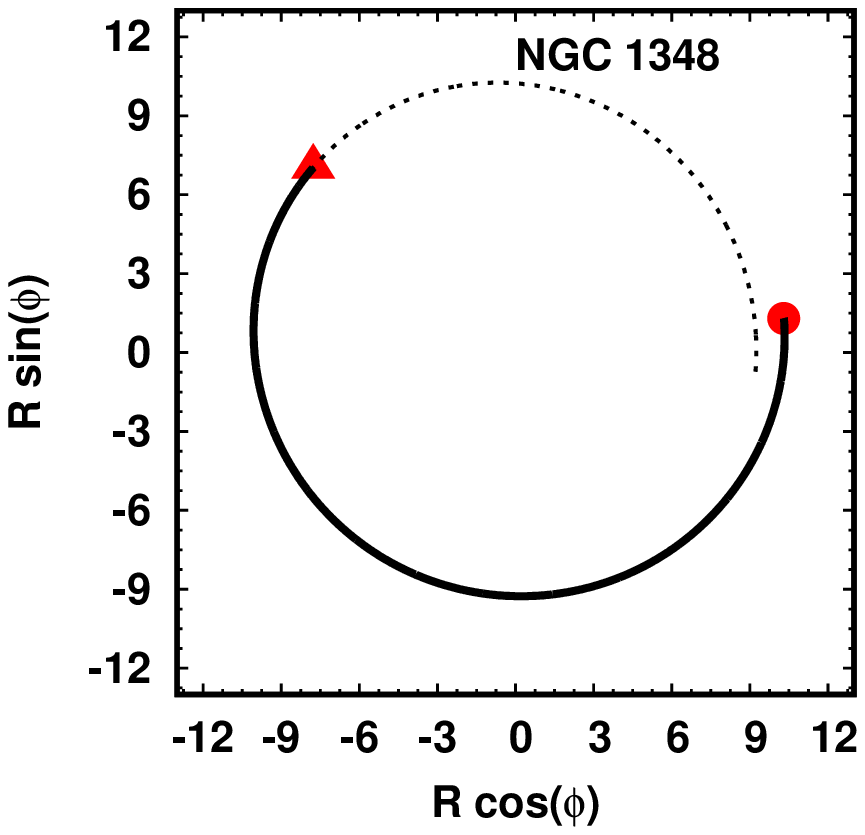}
}
\hspace{-4cm}\includegraphics[width=6.2cm, height=6.2cm]{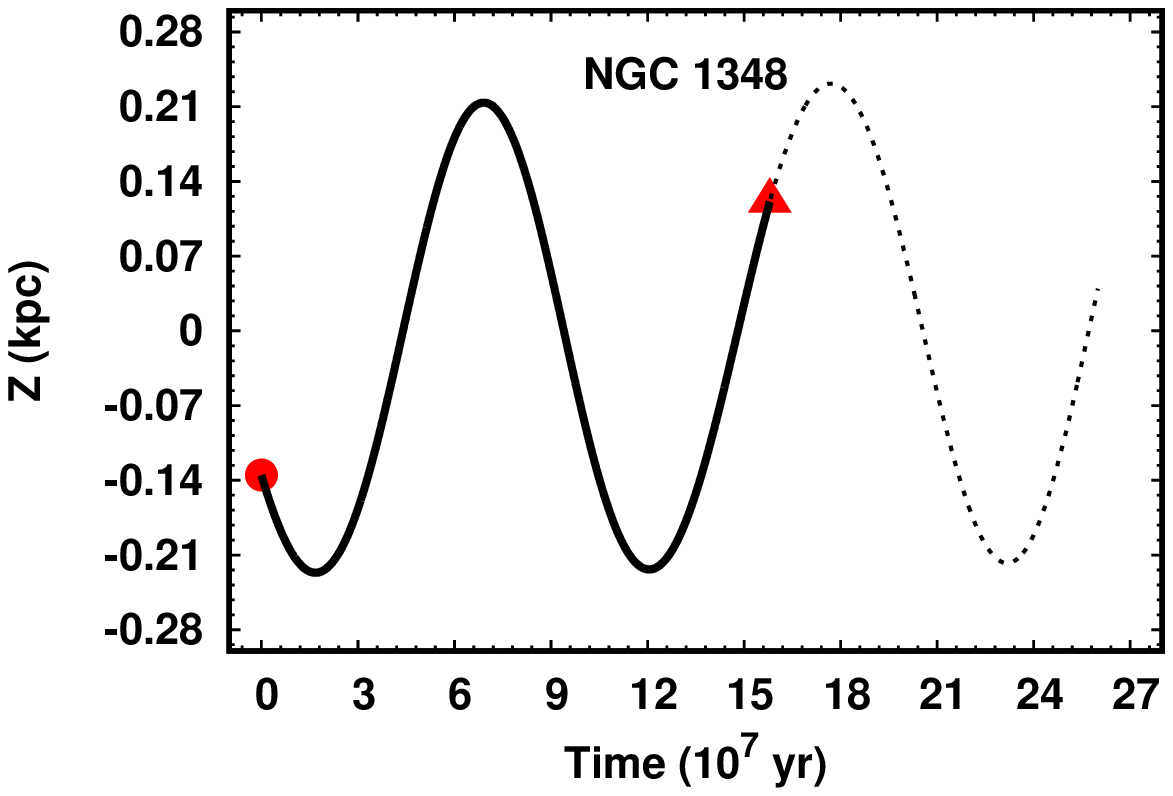}

\caption{Galactic orbits of the cluster NGC 1348 estimated with the Galactic potential model described in the text in the
time interval equal to the age of the cluster. The top left panel shows the side view and the top right panel shows the top
view of the orbit. Bottom panel shows the distance of cluster from the Galactic plane as a function of time. Dotted
line represents the cluster's orbits for a time interval of 284 Myr. The filled circle and the triangle sign denote the
birth and the present day position of cluster in the Galaxy.}
\label{orbit}
\end{center}
\end{figure*}

Fig.~\ref{orbit} shows orbits of the cluster  NGC 1348. In the top left panel, the motion of the cluster is described in terms of distance from
Galactic center and Galactic plane, which indicates the 2D side view of the orbit. In the top right panel, the cluster motion
projected into the plane of the Galaxy is described, which shows the top view of orbit. The bottom panel of this figure indicates
the distance of NGC 1348 from the Galactic plane as a function of time. 
The nearly circular orbit of NGC 1348 follows a boxy pattern. However the 
cluster shows a small drift of $\sim$ 83 pc from the circular motion. 
The birth and the present day position of NGC 1348 in the Galaxy
are represented by filled triangle and circle in Fig. \ref{orbit}. We also calculated the orbital parameters for the clusters which are
listed in Table \ref{orpara}. Here $e$ is eccentricity, $R_{a}$ is apogalactic distance, $R_{p}$ is perigalactic distance, $Z_{max}$
is the maximum distance traveled by cluster from Galactic disc, $E$ is the average energy of orbits, $J_{z}$ is $z$ component of angular
momentum and $T$ is time period of the revolution around the Galactic center. 
The orbital parameters determined in the present analysis are similar to the
parameters determined by Wu et al. (2009).

\begin{table*}
   \caption{Position and velocity components in Galactocentric coordinate system. Here $R$ is the Galactocentric distance,
     $Z$ is vertical distance from the Galactic disc, $U$ $V$ $W$ are radial tangential and vertical components of velocity
     respectively, and $\phi$ is the position angle relative to the sun's direction.
}
   \vspace{1cm}
   \centering
   \begin{tabular}{ccccccccc}
   \hline\hline
   Cluster   & $R$ &  $Z$ &  $U$  & $V$  & $W$ & $\phi$   \\
   & (kpc) & (kpc) & (km/sec) &  (km/sec) & (km/sec) & (radians)    \\
  \hline
   NGC 1348 & 10.39 & -0.14 & $-11.71 \pm 1.68 $  & $-232.55 \pm 1.51 $ &  $-10.14 \pm 1.65$  &  0.13    \\
\hline
  \end{tabular}
  \label{inp}
  \end{table*}
\begin{table*}
   \caption{The obtained orbital parameters using Galactic potential model.
   }
   \vspace{1cm}
  \centering
   \begin{tabular}{ccccccccc}
   \hline\hline
   Cluster  & $e$  & $R_{a}$  & $R_{p}$ & $Z_{max}$ & Birth position & $E$ & $J_{z}$ & $T$   \\
           &    & (kpc) & (kpc) & (kpc) & (R,Z) & $(100 km/sec)^{2}$ & (100 kpc km/s) & (Myr) \\ 
   \hline\hline
   NGC 1348 & 0.004 & 10.47  &  10.38  & 0.25 & (11.47,0.30) & -9.88 & -24.38  & 284 \\
 \hline
  \end{tabular}
  \label{orpara}
  \end{table*}

\section{Dynamics of the cluster}

\subsection{Luminosity and mass function}

\begin{figure}
\begin{center}
\hbox{
\includegraphics[width=8.2cm, height=8.2cm]{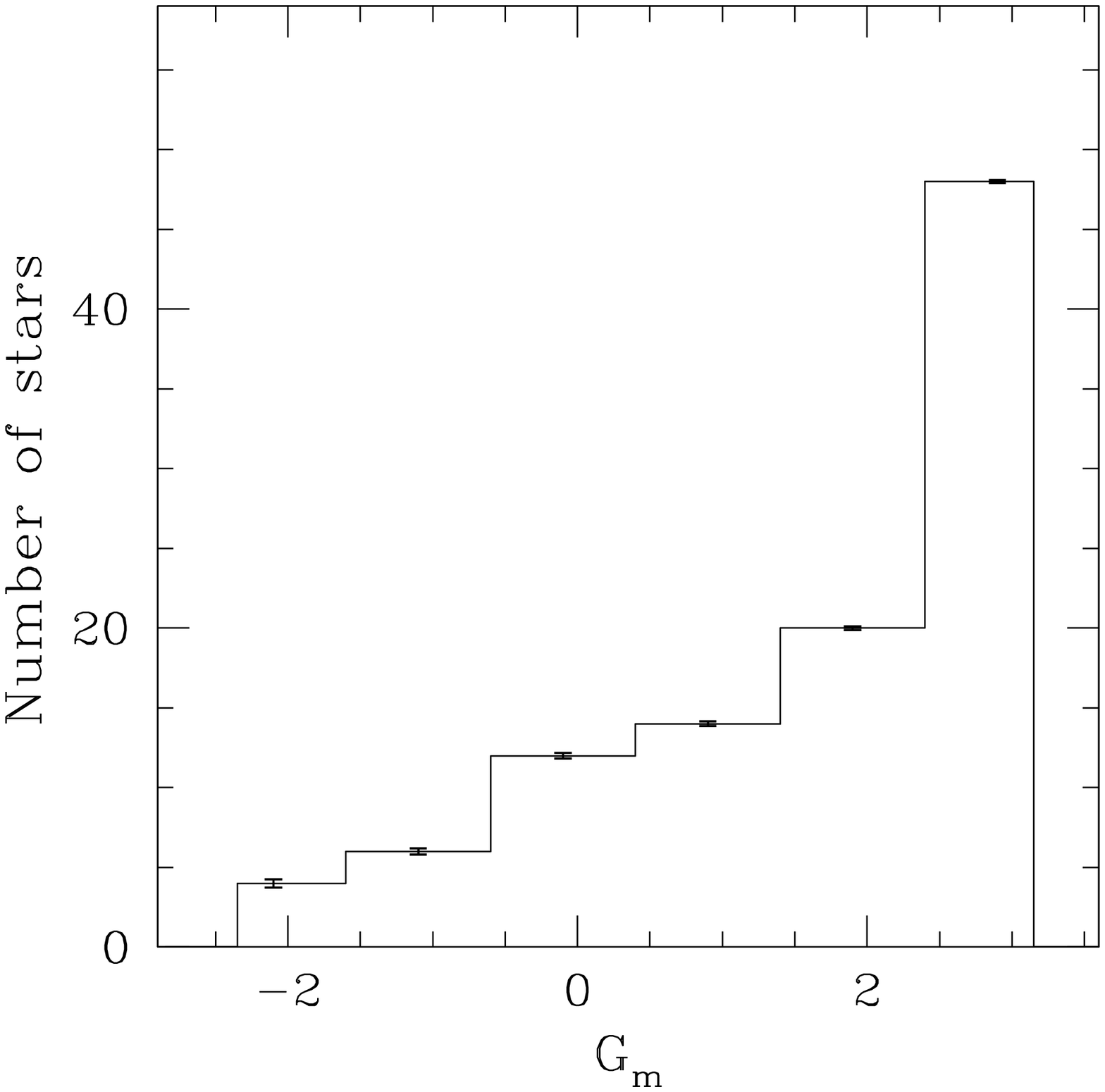}
\includegraphics[width=8.2cm, height=8.2cm]{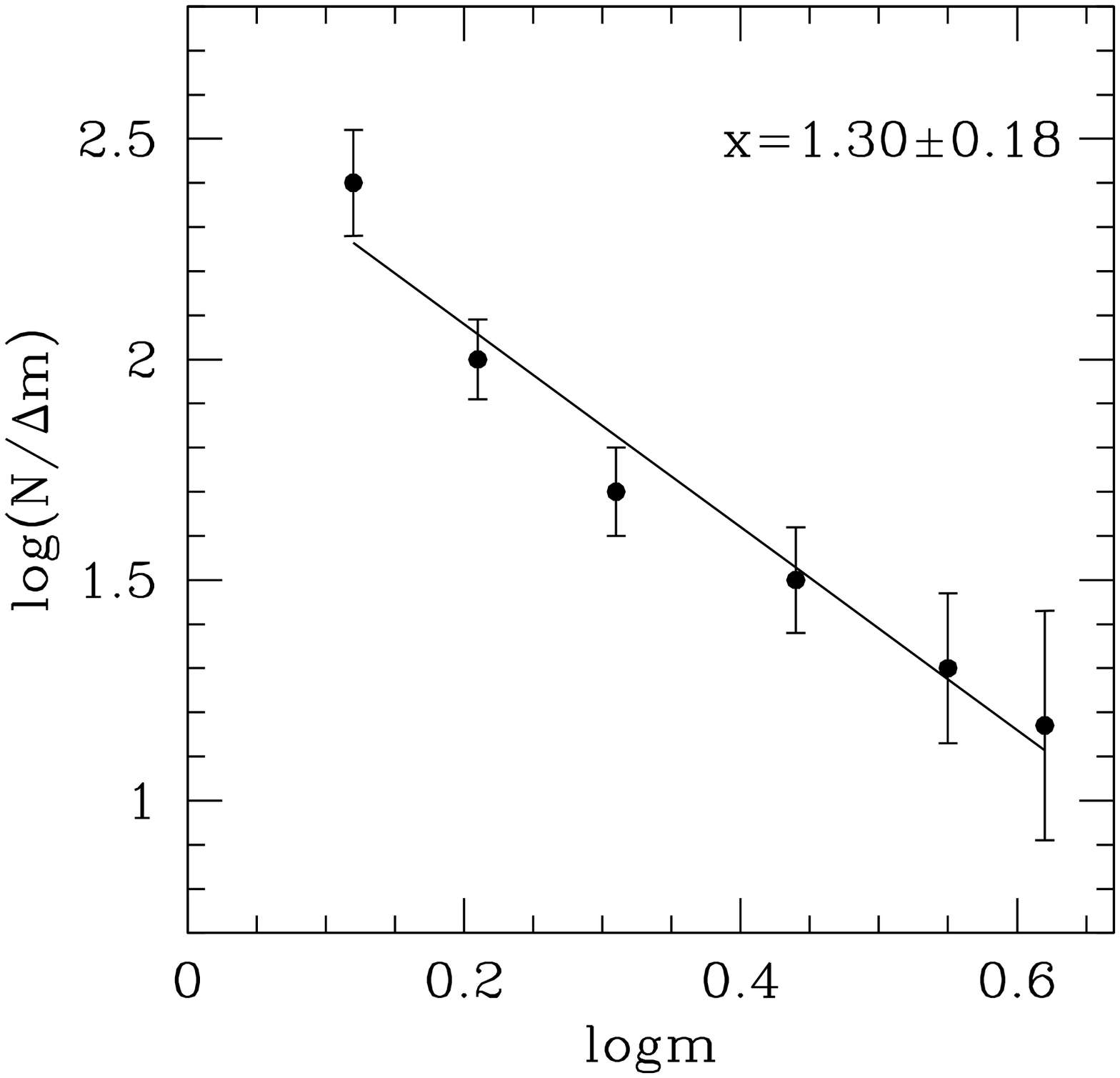}
}
\caption{(Left panel) Luminosity function of stars in the region of the cluster NGC 1348. (Right panel) Mass function
derived using the most probable members, where solid line indicates the power law given by Salpeter (1955).
The error bars represent $\frac{1}{\sqrt{N}}$.}
\label{lf_mass}
\end{center}
\end{figure}

Distribution of the cluster members in a unit magnitude range is called the 
luminosity function (LF). To derive the LF, we used only the probable members 
of NGC 1348. To construct the LF, we converted the apparent $G$ magnitudes 
into the absolute ones using the distance modulus. 
The resulting histogram is shown in the left panel of Fig. \ref{lf_mass}. 
This figure shows that the LF continues rising up to $M_{G}\sim$ 2.9 mag.

The LF and mass function (MF) are associated with each other according to the mass-luminosity relation (MLR).
We have used the theoretical tables of evolutionary tracks of Marigo et al. (2017) to convert luminosities into masses.
Fig. \ref{lf_mass} displays the luminosity function of member stars of the cluster (left panel) and the derived
present day mass function (PDMF; right panel). The PDMF, under specific conditions is an approximate representation of the IMF.

The shape of the present day mass function of stars in NGC 1384 for masses $\ge$ 1 $M_{sol}$ can be approximated by a
power law of the form,\\

\begin{equation}
~~~~~~~~~~~~~~~~~~~~~~~~~~~~~~~\log\frac{dN}{dM}=-(1+x)\log(M)+constant\\
\end{equation}

Where dN is the number of stars in the mass interval M+dM. We derive a value of $x=1.30\pm0.18$, which is close to the value of 1.35
derived by Salpeter (1955) for the nearby Galactic field. It is worth mentioning that the slope of the mass function of the Galactic
field is being constantly updated using more modern data and sophisticated inference techniques. A recent work by Mor et al. (2019)
inferred a shallower than Salpeter slope for the Galactic IMF (close to -1) which is in agreement with the theoretical prediction
of Dib \& Basu (2018). Dib et al. (2017) inferred the distribution function of the slope of the IMF for a large population of Galactic
clusters and found that it is well represented by a Gaussian distribution centered around the Salpeter value but with a standard deviation
of ~0.6. For NGC 1348, our derived value falls well within this range, when considering the uncertainty we have measured 
for the slope. The total mass was obtained as $\sim$215 $M_{sol}$.

\subsection{Mass-segregation study}

\begin{figure}
\begin{center}
\includegraphics[width=8.2cm, height=8.2cm]{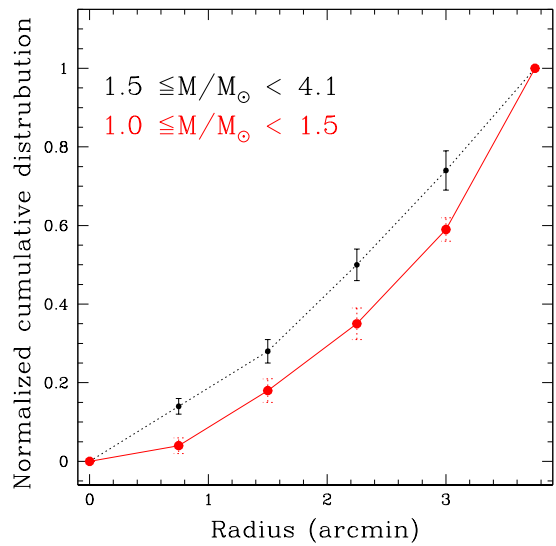}
\caption{The cumulative radial distribution for NGC 1348 using the most probable members in several mass ranges.} 
\label{mass_seg} 
\end{center}
\end{figure}

The mass segregation effect in the clusters has been described by many authors 
(e.g. Sagar et al. 1988; Hillenbrand \& Hartmann 1998; Fisher et al. 1998; 
Meylan 2000; Baumgardt \& Makino 2003; Dib, Schmeja \& Parker 2018;
Dib \& Henning 2019; Alcock \& Parker 2019). 
To understand this effect in NGC 1348, we divided 
the mass range in two subranges as 1.5$\le\frac{M}{M_{\odot}}\le$~4.1 and
1$\le\frac{M}{M_{\odot}}\le$~1.5. The cumulative radial stellar distribution of stars for two different mass ranges as
shown in Fig. \ref{mass_seg}. This figure demonstrates the mass-segregation effect as bright stars appear to be more centrally
concentrated than the low mass members. This has been checked through Kolmogrov-Smirnov test $(K-S)$. In this way, we found
that the confidence label of the mass-segregation effect is 91 $\%$.

The possible reason of the mass-segregation effect generally differs from one cluster to another. This may be because of dynamical
evolution or could be an imprint of star formation or both 
(Dib, Kim \& Shadmehri 2007; Allison et al. 2009; Pavlik 2020).
The most important result of this process is that the most massive stars sink gradually towards the cluster center and
transfer their kinetic energy to the more numerous lower-mass stars, thus leading to mass segregation. The relaxation time $T_{E}$
is defined as the time in which the stellar velocity distribution becomes Maxwellian and expressed by the following formula:\\

\begin{equation}
~~~~~~~~~~~~~~~~~~~~~~~~~~~~~~~~~~~T_{ES}=\frac{8.9\times10^5\sqrt{N}\times{R_{h}}^{3/2}}{\sqrt{\bar{m}}\times log(0.4N)}\\
\end{equation}

where $N$ represents the number of stars in the clusters (in our case the ones with membership probability higher than 50$\%$),
$R_{h}$ is the cluster half mass radius expressed in parsec and $\bar{m}$ is the average mass of the cluster members
(Spitzer \& Hart 1971) in the solar unit. The value of $\bar{m}$ is found as 2.07 $M_{\odot}$. The value of $R_{h}$ is assumed to
be equal to half of the cluster's extent. Using the above formula, the value of dynamical relaxation time $T_{ES}$ is determined
as 18 Myr\footnote{This value is obtained using stars  with mass $\ge$ 1 $M_{\odot}$. If we include the low
mass stars ($\ge$ 0.1 $M_{\odot}$), the value of the relaxation time becomes $\sim$ 52 Myr. In either case, the cluster is dynamically relaxed according
to this study.}. Hence, we conclude that NGC 1348 is a dynamically relaxed cluster.

\section{Kinematical structure of NGC 1348}

$\bullet$ {\textit{\textbf{Vertex (apex position) of the cluster}}}

The Apex coordinates are obtained by solving the geometric problems in which the intersection of vectors of spatial velocities
(i.e. $V_{x}$, $V_{y}$, $V_{z}$) of stars on the celestial sphere, when the beginning of the vectors is moved to the point
of observations. A formal description of the method, diagramming technique, and formulas to determine the error ellipses can be found
in Chupina et al. (2001, 2006). This method has been used previously by some of us (Vereshchagin et al. 2014, Elsanhoury et al. 2018,
Elsanhoury 2020a, 2020b, Postnikova et al. 2020). 
The equatorial coordinates of the convergent point have the following forms:\\
i.e.

\begin{equation}
~~~~~~~~~~~~~~~~~~~A_{\circ}=\tan^{-1}\Big[\frac{\overline{V_y}}{\overline{V_x}}\Big].
\end{equation}

\begin{equation}
~~~~~~~~~~~~~~~~~~~~D_{\circ}=\tan^{-1}\Big[\frac{\overline{V_z}}{\sqrt{\overline{V_x}^2+\overline{V_y}^2}}\Big].
\end{equation}

\begin{figure}
\centering
\includegraphics[width=10.5cm,height=8.5cm]{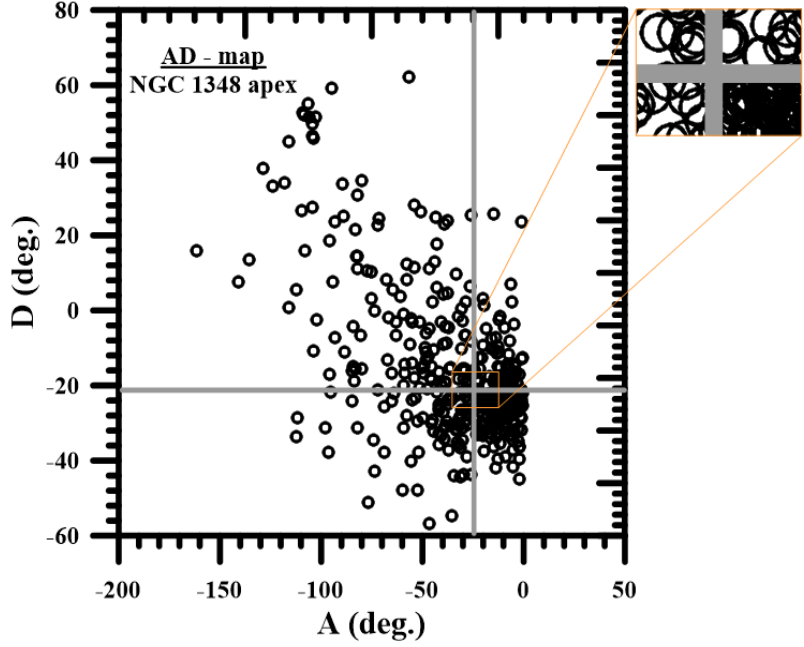}
\caption{The convergent point of NGC 1348 open cluster with AD-chart method, showing the apex coordinates $(A_{o},
D_{o})= (13.865\pm0.238, -48.348\pm0.144)$.}
\label{cp1} 
\end{figure}

The apex equatorial coordinates for NGC 1348 are presented here with Fig. \ref{cp1}.

We have also derived several kinematical parameters, 
for example, the matrix elements $(\mu_{ij})$, 
direction cosines $(l_{j}, m_{j}, n_{j})$ etc. 
using techniques described by Bisht et al. (2020).
All these parameters are listed in Table~\ref{all}.

\begin{table*}
\small
\caption{Dynamical and kinematical parameters of NGC 1348.}
\begin{tabular}{lll}
\hline
Parameters & Numerical values & Reference  \\ \hline
No. of members (N) & 438 &Present study   \\
Age (log)        &8.20 &Present study \\
$(A_{o}, D_{o})$& $-23.815\pm0.135$, $-22.228\pm0.105$
&Present study \\
Cluster radius (arcmin)   & 7.5 & Present study \\
Cluster radius (pc)   & 5.67 & Present study \\
$T_{ES}$(Myr)   &18.00 & Present study \\
$\tau$ & 8.886 $\pm$ 0.413 & Present study\\

$(\overline{V_{x}}, \overline{V_{y}}, \overline{V_{z}})$, (km s\textsuperscript{-1})       & $94.40\pm0.10$, $-41.63\pm0.15$, $-42.13\pm0.15$ & Present study \\

$(\overline{V_{\alpha}}, \overline{V_{\delta}}, \overline{V_{t}})$, (km s\textsuperscript{-1})       & $-100.57\pm0.10$, $-43.95\pm0.15$, $153.78\pm0.09$ & Present study \\
($\lambda_{1}$, $\lambda_{2}$, $\lambda_{3}$) (km s\textsuperscript{-1})        &       5765040, 19837.3, 348.674     &       Present study   \\

($\sigma_{1}$, $\sigma_{2}$, $\sigma_{3}$) (km s\textsuperscript{-1})   &       2401.05, 140.845, 18.673     &       Present study   \\

$(l_{1}, m_{1}, n_{1})$\textsuperscript{o} &       0.339, 0.404, $-0.850$     &       Present study   \\

$(l_{2}, m_{2}, n_{2})$\textsuperscript{o} &       $-0.431$, $-0.737$, $-0.522$           &       Present study   \\

$(l_{3}, m_{3}, n_{3})$\textsuperscript{o} &       0.837, $-0.543$, 0.076           &       Present study   \\

$(x_{c}, y_{c}, z_{c})$ (kpc)      &       $-8.811$, $-11.911$, $-18.486$     &               Present study   \\

$B_{j}$, j=1, 2, 3   &       $-58\textsuperscript{o}.197$, $-31\textsuperscript{o}.436$, $4\textsuperscript{o}.340$       &       Present study   \\
$L_{j}$, j=1, 2, 3   &       $-50\textsuperscript{o}.030$, $120\textsuperscript{o}.283$, $-147\textsuperscript{o}.059$    &       Present study   \\

$X_{\odot}$ (kpc)  &       $-2.175\pm0.047$&       Present study   \\
        &       -1.933  &       Cantat-Gaudin \textit{et al.} (2020)    \\

$Y_{\odot}$ (kpc)  &       $1.414\pm0.038$&        Present study   \\
        &       1.2568  &       Cantat-Gaudin \textit{et al.} (2020)    \\

$Z_{\odot}$ (kpc)  &       $-0.168\pm0.013$&       Present study   \\
        &       -0.1495 &       Cantat-Gaudin \textit{et al.} (2020)    \\

$R_{gc}$ (kpc)       &       $10.476\pm0.102$&       Present study   \\
        &       10.349  &       Cantat-Gaudin \textit{et al.} (2020)    \\
$S_{\odot}$ (km/s) &       111.36 & Present study   \\
($l_{A}, \alpha_{A})_{w.s.v.c.}$    &       $-32.21$, $56.58$   &       Present study   \\
($b_{A}, \delta_{A})_{w.s.v.c.}$    &       $-23.82$, $22.23$   &       Present study   \\

 \hline
\end{tabular}
\label{all}
\end{table*}

\section{Conclusions}
\label{con}

We conducted an exhaustive photometric and kinematical study of the poorly studied northern open cluster NGC 1348
using UKIDSS, WISE, APASS, Pan-STARRS1 and Gaia~EDR3 data sets. We calculated the membership probabilities of the stars in
NGC 1348 and hence found 438 member stars with membership probabilities higher than $50\%$ and G$\le$20 mag.
To derive the fundamental parameters of the cluster, 
we used only these selected member stars. We also shed some light on
the dynamical and kinematical properties of the cluster. Our main findings are summarized follows:

\begin{itemize}

\item The cluster's center is obtained as: $\alpha = 53.51\pm0.03$ deg ($3^{h} 34^{m} 2.3^{s}$)
      and $\delta = 51.41\pm0.02$ deg ($51^{\circ} 24^{\prime} 36^{\prime\prime}$) with the help of the most
      probable cluster members. The radius of the cluster is determined as 7.5 arcmin using a
      radial density profile.\

\item Based on the vector point diagram and membership probability estimation of stars, we identified 438 most
      probable cluster members for this object. The mean PMs of the cluster is estimated as $1.27\pm0.001$ and
      $-0.73\pm0.002$ mas yr$^{-1}$ in both the RA and DEC directions respectively.\

\item The distance is determined as $2.6\pm0.05$ kpc. This value is in fair agreement with the distance estimated
      using the mean parallax of the cluster. 
Age is determined as $160\pm40$ Myr by comparing the cluster's CMD with
      the theoretical isochrones given by Marigo et al. (2017).\

\item  The mass function slope is estimated as $1.30\pm0.18$, which is in good agreement with the value (1.35) given
       by Salpeter (1955) for field stars in Solar neighborhood.\

\item  Mass segregation is also observed for NGC 1348. The K-S test indicates $91\%$ confidence level of
       the mass-segregation effect. Our study indicates that NGC 1348 is a dynamically relaxed open cluster.\

\item The Galactic orbits and orbital parameters were estimated using Galactic potential models. We found that
      NGC 1348 is orbiting in a boxy pattern.

\item The apex position $(A, D)$ is computed with the AD-chart methods as: 
$(A_\circ, D_\circ)$ = (-23$^{\textrm{o}}$.815 $\pm$ 0$^{\textrm{o}}$.135, $-$22$^{\textrm{o}}$.228 $\pm$ 0$^{\textrm{o}}$.105) respectively.\

\item We computed the direction cosines ($l_{j}, m_{j}, n_{j}$) in three axes.\

\item The projected distance $(X_{\odot}, Y_{\odot}, Z_{\odot}$ are computed as ($-$2.175 $\pm$ 0.047, 1.414 $\pm$ 0.038,
$-$0.168 $\pm$ 0.013) kpc and the Solar elements $(S_\odot, l_A, b_A)$ are derived as $(111.36, -32^{\textrm{o}}.21, 56^{\textrm{o}}.58)$.

\end{itemize}

{\bf ACKNOWLEDGMENTS}\\

The authors thank the anonymous referee for the useful comments that improved the scientific content of the article
significantly. This work has been financially supported by the Natural Science Foundation of China (NSFC-11590782, NSFC-11421303).
Devesh P. Sariya and Ing-Guey Jiang are supported by the grant from the Ministry of Science and Technology (MOST),
Taiwan. The grant numbers are MOST 105-2119-M-007 -029 -MY3 and MOST 106-2112-M-007 -006 -MY3. This work has made use of
data from the European Space Agency (ESA) mission GAIA processed by Gaia Data processing  and Analysis Consortium (DPAC),
(https://www.cosmos.esa.int/web/gaia/dpac/consortium).



\newpage
{\bf appendix}

$\bullet$ {\textit{\textbf{Galactic longitude of the vertex $l_{2}$}}}

Stars in the thick disk ($\sim$ 200-300 pc) of our Galaxy 
appear to have formed well after the formation of the spheroidal
component of the Galaxy. 
As a result of nucleosynthesis in the stars, 
the distribution of velocity ellipsoids of the stars
in the Galactic plane, will be expected to have one axis 
directed exactly towards the Galactic center; 
this is known as the longitude of the vertex ($l_{2}$, Mihalas and Binney 1981). 
Our analysis of an ellipsoidal velocity distribution 
(Elsanhoury 2015 and 2020; Elsanhoury et al. 2016, 2018; \& Bisht et al. 2020) 
confirmed that the longitude of the vertex often differs significantly from zero 
i.e. $l_{2}$=$-$0.728 and gets affected by the stellar spectral classes (i.e. temperature scale).

$\bullet$ {\textit{\textbf{The Galactic longitude and Galactic latitude parameters}}}

Let $(L_{i})$ and $(B_{i})$, ($\forall$j=1, 2, 3) be the Galactic longitude and the Galactic latitude of the
directions, respectively which correspond to the extreme values of the dispersion, then
\begin{equation}
~~~~~~~~~~~~~~~~~~~L_{j}=tan^{-1}\Big(\frac{-m_{j}}{l_{j}}\Big),
\end{equation}
\begin{equation}
~~~~~~~~~~~~~~~~~~~B_{j}=sin^{-1}\Big(n_{j}\Big).
\end{equation}

$\bullet$ \textbf{\textit{The center of the cluster}}

The center of the cluster $(x_{c}, y_{c}, z_{c})$ can be derived by the simple method of finding
the equatorial coordinates of the center of mass for the number $(N_{i})$ of discrete objects,\\
i.e.
\begin{equation}
~~~~~~~~~~~~~~~~~~~~x_c=\left[\sum\limits_{i=1}^{N}d_i\cos\alpha_i\cos\delta_i\right]\Bigg{/}N,
\end{equation}
\begin{equation}
~~~~~~~~~~~~~~~~~~~~y_c=\left[\sum\limits_{i=1}^{N}d_i\sin\alpha_i\cos\delta_i\right]\Bigg{/}N,
\end{equation}
\begin{equation}
~~~~~~~~~~~~~~~~~~~~z_c=\left[\sum\limits_{i=1}^{N}d_i\sin\delta_i\right]\Bigg{/}N.
\end{equation}

$\bullet$ \textbf{\textit{Projected distances}}

Considering our estimated distances d(pc), we can calculate the distances to 
the Galactic center $(R_{gc})$ (Mihalas $\&$ Binney 1981) as a function of the Sun's distance from the Galactic center
(i.e. $R_{o}$=8.20$\pm$0.10 kpc) 
as mentioned recently with Bland-Hawthorn et al.(2019) as
R$_{gc}^{2}$=R$_{o}^{2}$+$d^{2}$-2$R_{o}$d$\cos{l}$. 
The projected distances towards the Galactic plane $(X_{\odot}$, $Y_{\odot})$ and
the distance from the Galactic plane $(Z_{\odot})$ (Tadross 2011) are computed as:
\begin{equation}
~~~~~~~~~~~~~~~~~~~~~~X_{\odot}=d\cos{b}\cos{l},
\end{equation}
\begin{equation}
~~~~~~~~~~~~~~~~~~~~~~Y_{\odot}=d\cos{b}\sin{l},
\end{equation}
\begin{equation}
~~~~~~~~~~~~~~~~~~~~~~Z_{\odot}=d\sin{b}.
\end{equation}

$\bullet$ \textbf{\textit{Solar elements}}

Let us consider a group with spatial velocities 
($\overline{\textrm{U}}$, $\overline{\textrm{V}}$ and $\overline{\textrm{W}}$). 
The components of the Sun's velocities 
($\textrm{U}_{\odot}$, $\textrm{V}_{\odot}$, and $\textrm{W}_{\odot}$) are then given as:
($\textrm{U}_{\odot}$=$-\overline{\textrm{U}}$), ($\textrm{V}_{\odot}$=$-\overline{\textrm{V}}$), and
($\textrm{W}_{\odot}$=$-\overline{\textrm{W}}$). Therefore, we have the Solar elements With spatial velocities considered (w.s.v.c.)
like;
\begin{equation}
~~~~~~~~~~~~~~~~~~~~~~~~S_{\odot}=\sqrt{\overline{U}^2+\overline{V}^2+\overline{W}^2},
\end{equation}
\begin{equation}
~~~~~~~~~~~~~~~~~~~~~~~~l_{A}=tan^{-1}\Bigg(\frac{-\overline{V}}{\overline{U}}\Bigg),
\end{equation}
\begin{equation}
~~~~~~~~~~~~~~~~~~~~~~~~~b_{A}=sin^{-1}\Bigg(\frac{-\overline{W}}{S_{\odot}}\Bigg).
\end{equation}

Now, let us consider the positions along the x, y, and z-axes 
in the coordinate system
which is centered at the Sun. Then, the Sun's velocities with
respect to this same group and referred to the same axes are given as; ($\textrm{X}_{\odot}^{\bullet}$=$-\overline{\textrm{V}}_{\textrm{x}}$).
($\textrm{Y}_{\odot}^{\bullet}$=$-\overline{\textrm{V}}_{\textrm{y}}$), and ($\textrm{Z}_{\odot}^{\bullet}$=$-\overline{\textrm{V}}_{\textrm{z}}$).
Therefore, we have obtained the Solar elements with radial velocities considered as;
\begin{equation}
~~~~~~~~~~~~~~~~~~~~~~~~~~S_{\odot}=\sqrt{\left(X_{\odot}^{\bullet}\right)^2+\left(Y_{\odot}^{\bullet}\right)^2+\left(Z_{\odot}^{\bullet}\right)^2},
\end{equation}
\begin{equation}
~~~~~~~~~~~~~~~~~~~~~~~~~\alpha_{A}=tan^{-1}\Bigg(\frac{Y_{\odot}^{\bullet}}{X_{\odot}^{\bullet}}\Bigg),
\end{equation}
\begin{equation}
~~~~~~~~~~~~~~~~~~~~~~~~~\delta_{A}=tan^{-1}\Bigg(\frac{Z_{\odot}^{\bullet}}{\sqrt{\left(X_{\odot}^{\bullet}\right)^2+\left(Y_{\odot}^{\bullet}\right)^2}}\Bigg).
\end{equation}

where ($l_{A}$, $\alpha_{A})$ is the Galactic longitude and right ascension of the Solar apex and
($b_{A}$, $\delta_{A})$ are the Galactic latitude and declination of the Solar apex. 
$(S_{\odot})$ is considered as the absolute value of the Sun's velocity 
relative to the stellar groups under investigation. \\

\end{document}